\documentclass[12pt]{article}       
\usepackage{amssymb}
\begin{document} 
\begin{center}
          {\large \bf The black disk to be observed in the Orear region} 

\vspace{0.5cm}                   
{\bf I.M. Dremin}

\vspace{0.5cm}              
          Lebedev Physical Institute, Moscow 119991, Russia

\end{center}
\begin{abstract}
It is argued that the very first signatures of the approach to the black disk
asymptotical limit in hadron collisions may be observed in the differential
cross section of elastic scattering. The exponentially decreasing with the
angle (or $\sqrt {\vert t\vert }$) regime beyond the diffraction peak will 
become replaced by an oscillatory behavior or by the power-like falloff. 
Some estimates of energies where this can happen are presented.
\end{abstract}

\medskip

Hadrons behave as extended objects in high-energy collisions. The extreme
model widely discussed for many years is the asymptotical picture of two
absolutely black Lorentz-contracted disks with complete absorption at any
impact parameter. In the impact parameter formalism the total cross section
is given by the expression
\begin{equation}
\sigma _t(s)=2\int d^2b(1-e^{-\Omega (s,\bf b)}).
\label{cstb}
\end{equation}
The integration is over the two-dimensional space of the impact parameter 
$\bf b$. $\Omega (s,\bf b)$ denotes the opaqueness (or blackness) 
of disks colliding with the c.m.s. energy $\sqrt s$ and the impact 
parameter $\bf b$. For absolutely black ($\Omega (s,{\bf b})\rightarrow \infty ,
\; b\leq R; \;\; \Omega (s,{\bf b})=0, \; b>R $) 
and logarithmically expanding ($R=R_0\ln s, \; R_0$=const ) disks with 
radii $R$ one gets from (\ref{cstb})
\begin{equation}
\sigma _t(s)=2\pi R^2+O(\ln s).
\label{cstR}
\end{equation}

The elastic and inelastic processes should contribute on equal footing
\begin{equation}
\frac {\sigma _{el}}{\sigma _t(s)}=
\frac {\sigma _{in}}{\sigma _t(s)}=\frac {1}{2}\mp O(\ln ^{-1}s).
\label{frcs}
\end{equation}

The width of the diffraction peak $B^{-1}(s)$ should shrink because
its slope increases as
\begin{equation}
B(s)=\frac {R^2}{4}+O(\ln s).
\label{wid}
\end{equation}

The forward ratio of the real part to the imaginary part of the elastic 
scattering amplitude $\rho _0$ must vanish asymptotically as
\begin{equation}
\rho _0(s)=\frac {\pi }{\ln s}+O(\ln ^{-2}s).
\label{rho0}
\end{equation}

None of these asymptotical predictions were yet observed in experiment.

Total cross sections increase with energy albeit all fits taking into account
the latest LHC data \cite{totem} prefer the power-like behavior. The
preasymptotic origin may be ascribed to this effect, especially in view
of some theoretical guesses \cite{cww} proposed long time ago.

Inelastic processes are about 3 to 4 times more abundant than elastic ones.
Even though the elastic cross section increased from 18$\%$ of the total cross 
section at ISR energies to 25$\%$ at LHC energies, it is much less than 50$\%$
shown above.

The diffraction peak shrinks about twice from energies
about $\sqrt s \approx 6$ GeV where $B \approx 10$ GeV$^{-2}$ to the LHC
energy where $B \approx 20$ GeV$^{-2}$. At ISR energies the slope $B(s)$
increased logarithmically. To account for LHC data one needs a stronger than
simple logarithmical dependence. The terms proportional
to $\ln ^2s$ are usually added in phenomenological fits. Even then the
predictions \cite {okor, caok} are not completely satisfactory. 
In connection with the power-like preasymptotical behavior of $\sigma _t$ one 
could expect at present energies the faster than logarithmical shrinkage of 
the diffraction peak as well. 

As a function of energy, the ratio $\rho _0$ increases from below, crosses 
zero and becomes positive at high energies. 
This is a general tendency for collisions of any initial
particles. For $pp$-scattering, the prediction of (\ref{rho0}) with 
values of $s$ scaled by 1 GeV  is still 
somewhat higher (about 0.177) than estimates 
from dispersion relations (0.14 in Refs \cite{drna, blha}) even at 7 TeV 
while strongly overshoots them at ISR where $\pi /\ln s \approx 0.37$.
No logarithmic decrease is seen in these predictions which, however, depend 
on behavior of the total cross section at higher energies. Moreover, the value 0.14
can only be reached according to (\ref{rho0}) at the energy 75 TeV. Probably,
at energies higher than 75 TeV the first signs of approach to asymptotics will 
become visible. No data about $\rho _0$ at the LHC energies exist yet. 

Thus we can only hope that we are now in the preasymptotical regime, at best.

Surely, there is another more realistic possibility that the black disk model 
is too extreme and the gray fringe always exists. It opens the way to the 
numerous speculations with many new parameters about the particle shape and 
opacity (see, e.g., \cite{cyan, bour, desg, isla, gbw, petr, bloc, kope, kovn,
trot, mesh} which, nevertheless, are not as extreme as the black disk model.

However, let us study just the black disk model. There exists another parameter 
which can shed some light on the situation. This is the ratio 
\begin{equation}
Z=\frac {4\pi B}{\sigma _t},
\label{zbs}
\end{equation}
which should be equal to 1/2 for black disks. In the Table 1 we show how this 
ratio evolves with energy. All entries there except the last two are 
taken from Ref. \cite{cyan} with simple recalculation $Z=1/4Y$. The data at 
Tevatron and LHC energies are taken from Refs \cite{am, totem}. All results are 
for $pp$-scattering except those at 546 and 1800 GeV which are for $p\bar p$
processes. The accuracy of the numbers listed in the Table 1 can be very
approximately estimated as $\pm 0.1$ from known error bars for the total cross
sections and the slopes. 

\medskip

Table 1

\medskip

\noindent $\sqrt s$, GeV| 2.70  4.11   4.74   6.27   6.98   7.62   13.8   62.5   546   1800   7000 \\
--------------------------------------------------------------------------------------- \\
$\;\;\;\;\;\;\; $  $Z$  $\;\;\;\;\;\;\;\;\; $      | 0.64  1.02  1.09  1.26  1.31  1.34  1.45  1.50  1.20  1.08  1.00 \\

We see that $Z$ increases from low energy region to ISR energies and then drops
down approaching 1 at LHC. Our interest in this ratio is related with the fact
that the differential cross section of elastic scattering beyond the diffraction
peak in the so-called Orear region is very sensitive to its value.

This region can be well fitted at different energies from rather low ones
\cite{adg} to TOTEM data at LHC \cite{dnec} by the very general expressions
describing the cross section falloff with the dip-bump substructure derived
in Refs \cite{anddre, anddre1}. It was shown there that the unitarity relation 
outside the diffraction peak can be reduced to the integral equation for the 
imaginary part of the amplitude. No assumptions have been made other than the 
validity of experimental data in the diffraction peak. Using the shape of the amplitude
in the diffraction cone
\begin{equation}
A_d(p,\theta )= 4ip^2\sigma _te^{-Bp^2\theta ^2/2}(1-i\rho _d),
\label{ampl}
\end{equation}
one gets the unitarity equation  written as
\begin{equation}
{\rm Im}A(p,\theta )=\frac {p\sigma _t}{4\pi \sqrt {2\pi B}}\int _{-\infty }
^{+\infty }d\theta _1 e^{-Bp^2(\theta -\theta _1)^2/2} (1+\rho _d\rho _l)
{\rm Im}A(p,\theta _1)+F(p,\theta ).
\label{linear}
\end{equation}
The amplitude is denoted by $A$. $p\approx \sqrt s/2$ and $\theta $ denote the momentum and the 
scattering angle in the center of mass system. $\rho _i$'s take into account 
the real parts at the corresponding angles in the diffraction cone ($i=d$)
and at larger angles ($i=l$). The integral term represents the two-particle 
intermediate states of the incoming particles. The function $F(p,\theta )$, 
called following Ref. \cite{hove} as the overlap function, represents the 
shadowing contribution of the inelastic processes to the elastic scattering 
amplitude.
 
Its analytic solution was obtained in a model-independent way with 
the assumptions that the role of the overlap function $F(p,\theta )$ is 
negligible outside the diffraction cone\footnote {The results of the papers 
\cite{dnec, ads} give strong support to this assumption. The overlap function
has been calculated there directly from experimental data subtracting the
elastic contribution $I_2$ from the lefthand side of the unitarity equation
wthout model assumptions. It is extremely small outside the diffraction peak,
and becomes even smaller at the LHC enargy compared to the lower ones.} and 
the real parts may be replaced by their average values in the diffraction peak 
$\rho _d$ and outside it $\rho _l$, correspondingly. Let us stress once more 
that the Gaussian shape of the amplitude has been only used at rather small 
angles in accordance with experimental data. If it is ascribed to the Pomeron
exchange, one is tempted to consider this solution as an effect of shadowing 
corrections due to multi-Pomeron graphs. Such corrections near the diffraction 
cone were considered, e.g., in Refs. \cite{kmr, glm}. Even though their 
conclusions differ, the qualitative tendency of peak suppression is valid. 
The definite model summing multi-Pomeron graphs was considered in \cite{adya}. 
However, the oscillations over the exponentially damped regime happened to
be too strong. The non-linear unitarity equation was solved by iterations 
assuming the Gaussian shape of the overlap function \cite{anddre1} but did 
not lead to quantitative agreement with experiment. That is ascribed to the
ad hoc chosen shape of the overlap function which was recognized as
unsatisfactory in earlier works (e.g., see \cite{fuku, zale, mich, giff}).
The solution of Eq. (\ref{linear}) does not depend on model assumptions.

Here we show only the leading term of the solution describing the exponential
falloff with the angle $\theta $ (or $\sqrt {\vert t\vert }$) within the Orear 
regime which is most important for our purpose:
\begin{equation}
{\rm Im} A(p,\theta )=C_0(p)\exp \left (-\sqrt 
{2B\ln \frac {Z}{1+\rho _d\rho _l}}p\theta \right ).
\label{sol}
\end{equation}
The solution predicts the dependence on $p\theta \approx \sqrt {\vert t\vert }$ 
but not the dependence on the collision energy. Its validity may be checked by
the direct substitution in Eq. (\ref{linear}). The values of the parameter $Z$
discussed above are crucial for the slope of the differential cross section in
the Orear region. Another important parameter is the product of the average 
values of ratios of real to imaginary parts of the amplitude in the diffraction
cone and in the Orear region. While the former ratio can be estimated from the
experimental results on interference of Coulomb and nuclear amplitudes 
supported by calculations with the help of the dispersion relations, the 
latter one was never estimated from experimental data. Any information about it 
becomes valuable. That is why it is important that from the fit 
\cite{dnec} at 7 TeV it became possible for the first time to get its
estimation from the experimentally measured slope in the Orear region. It 
happened to be surprisingly large in the absolute value ($\rho _l\approx -2$). 
The high sensitivity of the expression (\ref{sol}) to this parameter is 
determined by the value of $Z$ being practically equal to 1 as seen 
in the Table 1 and, correspondingly, of $\ln Z$ very near 0. 

Once the expression for the imaginary part of the amplitude (\ref{sol}) outside
the diffraction peak is known, it is possible to calculate the real part in
the same region assuming the geometrical scaling \cite{ddd} is valid.
This has been done in \cite{real}. It is shown that the real part changes its 
sign in accordance with the general theorem proven in \cite{mar2} and becomes
negative in the Orear region. Its absolute values happen, however, to be much
smaller than those required by the fit of TOTEM data at 7 TeV. 
This problem has not been resolved yet.

Now, the question arises about what happens if the black disk limit of $Z=1/2$
will be approached at higher energies. If the exponential slope persists,
the only solution would be to use even higher absolute values of $\rho _l$ 
(but negative!). However, this would show that there is no approach to black 
disks at all. This limit asks for $\rho _d\propto \ln ^{-1}s$ and,
consequently, to $\rho _l\propto \ln s$ that seems too exotic possibility.
If the product $\rho _d\rho _l$ tends to zero that looks more realistic,
then the behavior of the differential cross section in the Orear region
should become cardinally changed. In place of the exponentially decreasing
falloff, the oscillatory behavior should appear. Now, the solution of the 
equation (\ref{linear}) looks as
\begin{equation}
{\rm Im} A(p,\theta )=C_0(p)\cos 
\left (\sqrt {2B\ln \frac {1+\rho _d\rho _l}{Z}}p\theta \right ).
\label{sbl}
\end{equation}
The imaginary part of the amplitude dominates and determines the 
shape of the differential cross section
\begin{equation}
\frac {d\sigma}{dt}=\frac {C_0^2}{16\pi s^2}\cos ^2\left (\sqrt {2B\ln \frac {1+\rho _d\rho _l}{Z}\vert t\vert } \right ).
\label{dcsc}
\end{equation}
The oscillatory behavior with periodically placed zeros is clearly seen.

In the black disk limit $Z=1/2; \; \rho _i=0; \; B=R^2/4$. Thus
\begin{equation}
{\rm Im} A(p,\theta )=C_0(p)\cos (\sqrt {0.5R^2\ln 2\vert t\vert } )=
C_0(p)\cos (\sqrt {\sigma _t\ln 2/4\pi } p\theta ).
\label{Rbl}
\end{equation}
The distance between oscillation zeros becomes very small and asymptotically 
tends to zero
\begin{equation}
\Delta(\sqrt {\vert t\vert } )=\frac {\pi }{\sqrt {2B\ln \frac {1+\rho _d\rho _l}{Z}} }.
\label{dpt}
\end{equation}
It is hard to predict at what energies this transition from one regime to
another will happen. The only guess I can propose is to rely on possible
approach of the values of $\rho _d\approx \rho _0$ evaluated from dispersion 
relations to the black disk limit (\ref{rho0}). As discussed above, it can 
happen at energies about 75 TeV.

In some way, the conclusion about the undamped oscillations looks contradictory.
Hardly the differential cross section will not decrease with $\vert t\vert $
at all. It is more probable that the factor $C_0^2/s^2$ is so strong that the 
regime (\ref{dcsc}) dies out very fast with energy increase. Then it becomes 
unobservable and the place of the Orear region is filled by the hard parton
scattering. The exponential with $\vert t\vert ^{1/2}$ falloff will be 
replaced by the power-like decrease.

Anyway we conclude that the approach to the black disk limit will induce
drastic changes in the behavior of the differential cross section in the
Orear region replacing the exponential falloff either by undamped oscillations
in some energy interval or, that is more probable, by power-like behavior due 
to the hard parton scattering. It is possible that this feature will become 
the very first one noticeable if collision partners remind the black disks 
at extremely high energies.

\medskip

{\bf Acknowledgement}

This work was supported by the RFBR grant 12-02-91504-CERN-a and by the 
RAN-CERN program.

\end{document}